

%
\magnification=1200
\font\text=cmr10
\font\it=cmti10
\font\subtitle=cmr10 scaled \magstep2
\def\n{\noindent}
\font\caption=cmr9
\baselineskip=16pt
\text
\def\ga{\alpha} \def\gb{\beta} \def\g{\gamma}\def\gs{\sigma}
  \def\gq{\theta}
 \def\gl{\lambda} \def\gd{\delta} \def\gD{\Delta}
\def\p{\partial}
\def\half{{1\over 2}}
\def\gr{\rho} \def\SE{S_{\rm eff}}
\def \k{\kappa}
\def\vp{\varphi} \def\pn{\partial_{\mu}}
\def\pmu{\partial^{\mu}} \def\ps{\psi^{-1}}
\null

\rightline {SU-ITP-92-2}
\rightline {DAMTP-1-1992}
\rightline {January 1992}
\bigskip\bigskip
\centerline {\subtitle   Scalar -- Tensor Quantum Gravity}
\centerline {\subtitle  in Two Dimensions}

\vskip 1.4cm
\centerline {J. G. Russo\footnote {$^*$}{e-mail: russo@slacvm.bitnet.}}
\smallskip
\centerline {\it Theory Group, Department of Physics and SLAC}
\centerline {\it Stanford University, Stanford, CA 94305}
\bigskip
\centerline {A. A. Tseytlin\footnote {$^{**}$}
{e-mail: aat11@amtp.cam.ac.uk  .  On leave of
absence from the Department of Theoretical Physics, P.N. Lebedev
Physics Institute, Moscow 117924, Russia.  }}
\smallskip
\centerline{\it DAMTP, Cambridge University}
\centerline {\it Cambridge CB3 9EW, United Kingdom}
\vskip 1.8cm

{\caption  We discuss some classical and quantum properties of
2d gravity models involving  metric and a scalar field.
Different models are parametrized in terms of a scalar potential.
We show that a general Liouville - type model with exponential
potential and linear curvature coupling is renormalisable
at the quantum level while a particular
 model  (corresponding to $D=2$ graviton - dilaton
 string effective action  and having a black hole
solution ) is finite. We use the condition of a
``split'' Weyl symmetry  to  suggest possible  expressions
 for the ``effective action" which  includes the quantum
anomaly term.}
\vfill\eject \null
\centerline{\bf 1. Classical action and solutions}\bigskip
A study of 2d quantum gravitational models may teach us important
lessons about  quantum gravity in general.
For example, one would like to understand the effect of
quantum  corrections on the properties of  classical
(e.g. black hole) solutions. A particular ``string-inspired"
2d model which contains black hole solutions [1]  was recently
discussed in ref.[2]. In this paper we shall consider some effects
of quantum gravitational corrections on such type of scalar - tensor
2d models. We shall first make a field redefinition which
brings the action into a canonical form (with different models being
parametrised by a  scalar potential). This
form of the action  facilitates the study of renormalisation and
is useful in trying to determine  the structure of the ``effective action"
which accounts for the general covariance of the quantum theory,
i.e. is invariant under the background (``split") Weyl symmetry in the
conformal gauge [3]. We shall find that the ``string-inspired" model
corresponds to a Liouville model with an
exponential potential and is finite within the standard
loop expansion. We shall suggest  two
possible ansatze  for the  corresponding ``effective action"
based on two different choices of a metric
in terms of which the anomaly contribution is
constructed.

We shall consider the following model for a scalar field interacting
with gravity in two dimensions\footnote{$^1 $}{We shall assume that
the metric has Euclidean signature. Continuation to Minkowski signature
is straightforward. We have chosen the standard plus sign for the kinetic term
of $\phi$. When $F$ is non-zero this does not guarantee that $\phi$ is a
physical field. In fact, the scalar field mixes with the conformal factor
of the metric and the signature of the resulting kinetic matrix is  (+ --).
Since the euclidean path integral
is in any case unstable and should be defined by an analytic
continuation the choice of the minus sign in front of (1.1) is also
legitimate. }
$$S=\int d^2x \sqrt{g}\big[ {1\over 2}g^{\mu\nu}\p_\mu\phi
\p_\nu\phi+ F(\phi)R +{\cal U}(\phi)\big]\ . \eqno(1.1)$$
Here $\phi $ is dimensionless so that ${\cal U} =\mu\  \bar {\cal U},
\ [\mu]=cm^{-2}$.
Eq.(1.1) is the most general local reparametrisation invariant
action containing terms of dimension $\leq 2$ (an
arbitrary function $K(\phi)$ in the kinetic term of the scalar field
can be absorbed into a redefinition of $\phi $).

The action (1.1)
is of interest as a simple tractable model of 2d gravity:
the introduction of a scalar field is necessary in order to get
a non-trivial local reparametrisation invariant action in
two dimensions (similar models were previously discussed in refs.[4]).
Related actions appear from higher dimensional Einstein action after a
reduction to two dimensions.

Power counting implies that (1.1) should be renormalisable in a
generalized sense (i.e. assuming that the form of the scalar functions
$F$ and \
$\cal U$ may change under renormalization). To study some formal
properties of (1.1) it is useful to make a redefinition of $\phi $ and
a Weyl rescaling of the metric which effectively replace
 $F$ by a linear function.
Within the perturbation theory expansion we may ignore possible
singularities of such field transformation.
In general, one should take into account  the region of definition
of the scalar field and also orders of critical points of $F$ [5].
As a result, eq.(1.1) takes the form
$$S=\int d^2x \sqrt{\tilde g}\big( {1\over 2} \tilde
g^{\mu\nu}\p_\mu\varphi
\p_\nu\varphi+ {1\over 2}q\varphi \tilde R + V(\varphi)
\big) \ ,\eqno(1.2)$$
where $q={\rm const }, \ \tilde R=R(\tilde g)$ and
$${1\over 2}q\varphi=F(\phi )\ ,\ \
 g_{\mu\nu}=e^{2\gr}\tilde g_{\mu\nu}\ , \ \ \eqno(1.3)$$
$$ \gr (\phi )=
{F\over q^2}-{1\over 4} \int^{\phi}d\phi' \big({dF\over d\phi'}\big)
^{-1}\ ,\ \   V={\cal U} e^{2\gr} \ .\eqno(1.4)$$
Thus the class of models (1.1) is parametrized by one arbitrary
function $V$ of the scalar field.

To illustrate the transformation (1.3),(1.4) let us consider the
metric -- dilaton functional (string effective action)
 which generates the $\gs $-model Weyl anomaly
coefficients in the case of $D=2$ target space\footnote{$^2$}{ $c $ is
proportional  to an effective
central charge, $\ \alpha '=1 \ $; a
  numerical factor in front of the
action can be absorbed into a constant part of $\Phi$.
We have chosen the plus sign in front of (1.5) in order for
the transformation which puts it into the form (1.1),(1.2) to be real (e.g.
not changing the sign of the metric). If we start with the action (1.5)
with the minus sign we get (1.2) also with the minus sign. As we have noted,
there is no  $ a\  priori $ reason for a particular choice of sign from
the present point
of view of two dimensional theory. However, if one interprets (1.5) as
originating from a higher dimensional theory (in which ``transverse" gravitons
should have physical sign in the action) or
demands correspondence with the string effective action which also includes
physical
scalar field  not coupled to the curvature (tachyon) one should
 change the sign in (1.5). While this choice of overall sign is
not important at the classical level it becomes relevant once the quantum
corrections are included (see sect.3). Though we shall stick to the
choice of the plus signs in (1.5),(1.2) it is easy to modify
the results of sect.3  to the case of the minus sign: one should
change the overall signs in (3.1),(3.13)-(3.15),(3.18)-(3.20) as well
as the signs of the anomaly coefficients $A, \ B $ (the sign of the anomaly
term is invariant).}
$$S={1\over 8}\int d^2x \sqrt {g} e^{-2\Phi}\big(R+ 4(\nabla \Phi )^2
+c\big) \ .\eqno(1.5)$$
Eq.(1.5) can be represented in the form (1.1):
$$S=\int d^2x \sqrt{g} \big[ {1\over 2} g^{\mu\nu}\p_\mu\phi
\p_\nu\phi +{1\over 8}\phi^2 R+{1\over 8}c\phi^2 \big]\ ,
\ \ \  \phi\equiv e^{-\Phi} \ .\eqno(1.6)$$
By applying the transformation (1.3),(1.4) to (1.6)
$$q\varphi={1\over 4}\phi^2\ ,\ \ \ \gr ={1\over 8q^2}\phi^2
-\log \phi \ \ , \ \ \ g_{\mu\nu}= {e^{\vp /q}\over 4q\vp}
{\tilde g}_{\mu\nu} \ ,\eqno(1.7)$$
we conclude that (1.5) takes the form ($\mu = c/8$ )
$$S=\int d^2x \sqrt{\tilde g} \big[ {1\over 2} \tilde
g^{\mu\nu}\p_\mu\varphi
\p_\nu\varphi +{1\over 2}q\varphi \tilde R+\mu e^{\varphi / q} \big]
\ , \eqno(1.8)$$
or
 $$S= {q^2 }\int d^2x \sqrt{\tilde g} \big[ {1\over 2} \tilde
g^{\mu\nu}\p_\mu\tilde \varphi
\p_\nu \tilde \varphi +{1\over 2} {\tilde \varphi}
 \tilde R+\mu e^{\tilde \varphi } \big]
\ . $$
 It is interesting to note  that the simple
Liouville structure of the potential in (1.8) is due to a particular
relative coefficient of the dilaton and Einstein terms in (1.5)
(if the coefficient 4 is replaced by $\gamma $ one finds (1.2) with
$V=\mu \varphi^{1-{1\over 4}\gamma }e^{\varphi/q} $). This particular
structure of (1.5) is also responsible for the
existence of a black hole - type
classical solution [1].

The classical field equations which follow from (1.2) are
$$-\tilde\nabla^2\varphi +{1\over 2}q\tilde R+V'=0\ ,\eqno(1.9)$$
$$q\tilde\nabla_\mu\tilde\nabla_\nu\varphi-\p_\mu\varphi\p_\nu\varphi
+\tilde g_{\mu\nu} [{1\over 2}(\p \varphi)^2+V-q\tilde\nabla^2
\varphi ]=0 \ .\eqno(1.10)$$
The trace of (1.10) represents the classical Weyl anomaly of (1.2):
$$-q\tilde\nabla^2\varphi+2V=0 \ .\eqno(1.11)$$
Eqs. (1.9) and (1.11) imply that
$$\tilde R={2\over q^2}(2V-qV')\ .\eqno(1.12)$$
Since
$q\tilde\nabla_\mu\tilde\nabla_\nu\varphi-\p_\mu\varphi\p_\nu\varphi
\sim \tilde g_{\mu\nu}$ the metric which solves (1.10) always has
the Killing vector [5]
$\ \xi_\mu=\varepsilon_{\mu\nu} {\tilde \nabla}^{\nu} (e^{-\varphi/q})$. In
conformal coordinates where $\tilde g_{\mu\nu}=e^{2\gs}\gd_{\mu\nu}$
eqs. (1.9) and (1.11) reduce to
$$-\p^2\varphi -q\p^2\gs+V'e^{2\gs}=0\ ,\eqno(1.13)$$
$$-q\p^2\varphi +2Ve^{2\gs}=0\ .\eqno(1.14)$$
In the case of exponential potential $V=\mu e^{\varphi/q}$
corresponding to (1.5),(1.8) one finds that eqs. (1.10),(1.13) and (1.14)
have a spherically symmetric solution
$$2\gs +{1\over q}\varphi =k\ ,\ \ \ \ \varphi=ax^2+b\ ,
\ \ \ \ \eqno(1.15)$$
$$ \ a={\mu\over q} e^k\ ,\ \ \ a,b,k={\rm const}\ . $$
In terms of the original metric and dilaton which appear in (1.5) this
corresponds to the $D=2$ ``black hole" solution of ref.[1]
$$\Phi=\Phi_0 -{1\over 2}\log \varphi\ ,$$
$$ g_{\mu\nu}=e^{2\gr
+2\gs}\gd_{\mu\nu}={1\over \phi^2}e^{2\gs+{1\over q}\varphi}
\gd_{\mu\nu}=e^{2\Phi +k}\gd_{\mu\nu}={d\over\varphi} \gd_{\mu\nu}
\ ,\ \ \     d= {e^k}/4q \ ,\eqno(1.16)$$
or in Minkowski notation
$$ds^2={dx^+dx^-\over   b' -a'x^+x^-}\ ,\
\ \Phi={\Phi'}_0 -{1\over 2} \log (b' -a'x^+x^-)\ ,\
\ a'={c\over 4},\  b'={M\over \sqrt {c}}\ .\eqno(1.17)$$
We see that
(1.7)  transforms the regular solution (1.15) into
the singular one: the zero of $\varphi $ in (1.15) is
transformed into the singularity of the original fields in (1.16).
\bigskip
Let us note that it is possible to find an explicit solution for
arbitrary potential $V$. This is easy to do in terms of the
variables $(g_{\mu\nu}, \Phi )$ used in (1.5). Consider the
following generalization of (1.5)
$$S={1\over 8}\int d^2x \sqrt {g}  e^{-2\Phi}\big[R+4(\p \Phi )^2
+U(\Phi )\big] \ ,\eqno(1.18) $$
$$ U=c +c_1e^{2\Phi}+c_2e^{4\Phi}+... \equiv c + \bar U
\  \ . $$
The corresponding equations of motion can be represented in the form
$$R_{\mu\nu}+2\nabla_\mu \nabla_\nu \Phi+ F_1(\Phi )g_{\mu\nu}=0
\ ,\eqno(1.19)$$
$$-{1\over 2}\nabla^2 \Phi +(\p \Phi)^2+F_2(\Phi) =0\ ,
\ \ \ F_1\equiv -{1\over 4}U'\ ,\ \ F_2\equiv -{1\over 4}U\ .
\eqno(1.20)$$
In $D=2$ $\ \ \ R_{\mu\nu}={1\over 2}g_{\mu\nu}R\ $ so that $g_{\mu\nu}$ is
proportional to $\nabla_\mu\nabla_\nu \Phi$. Hence the solution has
the Killing vector $\xi_\mu=\varepsilon_{\mu\nu}\nabla^\nu\Phi$.
Fixing the coordinates so that
$$ds^2=f d\gq ^2 + f^{-1} dx^2\ ,\ \ \ f=f(x)\ ,\ \  \Phi=\Phi (x)
\ , \eqno(1.21)$$
with $\gq$ parametrizing the direction along the Killing vector
flow, we find that the $(xx)$ and $(\gq\gq )$ components of (1.19)
reduce to ($R=-f''$)
$$-{1\over 2}f^{-1}f''+2\Phi''+f^{-1}f'\Phi'+f^{-1}F_1=0\ ,\ \
-{1\over 2}ff''+ff'\Phi '+fF_1=0\ .\eqno(1.22)$$
As a consequence,  $\Phi'' (x)=0$, i.e.
$$\Phi =\Phi_0- bx\ ,\eqno(1.23)$$
so that
$$f(x)=ae^{-2bx}-{2\over b}e^{-2bx}\int^x dx' e^{2bx'}\hat F_2(x')
\ ,$$
$$ \hat F_2(x)\equiv -{1\over 4}U(\Phi_0-bx)
\equiv -{1\over 4}\big( c +\hat U(x)\big)\ , \eqno(1.24)$$
i.e.
$$f(x)={1\over 4}{c \over b^2}+a e^{-2bx}+{1\over 2b}e^{-2bx}
\int^x dx' e^{2bx'}\hat U(x')\ .\eqno(1.25)$$
The equivalent solution was found in [6].
For a large class of potentials it represents
a black hole - type configuration
(if $\hat U=0$ the coordinate transformation
$x^+x^-=\ga+\gb e^{2bx},\ x^+/x^-=e^{2t}$
brings the metric into the form (1.17)).
     In the context of string theory $\bar U$ in (1.18) represents
higher loop corrections to the dilaton potential. To the one-loop
order we find from (1.25) [6]
$$f(x)={1\over 4}{c\over b^2}+a e^{-2bx}+hxe^{-2bx} +O(e^{-4bx})\ ,
\ \ a={M\over \sqrt c },\ h={c_1\over 2b}\ .\eqno(1.26)$$
 The genus 1
correction dominates over the ``mass term" $ae^{-2bx}$
in the weak coupling region $x\to \infty$.
The curvature
$R=-e^{-2bx}(4b^2{M\over \sqrt{c}}-2c_1+2bc_1x)
+O(e^{-4bx})$ has very different behaviour as compared to the
tree-level solution (in particular, it changes
sign at some point).
\bigskip\bigskip
\centerline{\bf 2.  Perturbative renormalization}\bigskip
Let us now discuss the renormalization of the model (1.2).
Though there are no propagating degrees of freedom, there are
nontrivial ultra-violet divergences. On dimensional grounds the
counter-terms should have the structure
$$\gD S=\int d^2x \sqrt{\tilde g}\big [ K(\varphi ) \tilde
g^{\mu\nu}\p_\mu\varphi
\p_\nu\varphi+ P(\varphi )\tilde R + Q(\varphi)
\big ]\ , \eqno(2.1)$$
so that after a non-linear renormalization  of the scalar field
and the metric the renormalised action
 takes the same form as the original
action (1.2) with some new potential $\tilde V$.\footnote{$^3$}
{Equivalently, one may use the classical field equations (1.9),(1.10)
to transform (2.1) into the form $\int d^2x\sqrt{\tilde g}\bar V
(\varphi )$.}
For a particular form of the potential
the model (1.2) is not renormalizable
in the usual sense unless $\tilde V$ is simply proportional to $V$.
It is straightforward to compute $\tilde V$ in the one-loop
approximation. The  ``on-shell" counter-term should
be gauge-independent so that we may use the simplest quantum conformal
gauge\footnote{$^4$}{The 1-loop renormalization of (1.1) in
harmonic-type gauges was recently discussed in ref.[7]. A surprising
result that the ``on-shell" counter-term is gauge dependent
was found. This is probably due to the fact that the boundary
divergences (which can mix with the volume ones on the equations
of motion (1.9),(1.10))
were ignored. Our conformal gauge result disagrees
with the expressions found in [7].},  $\ \tilde g_{\mu\nu}=
e^{2\gs }\bar g_{\mu\nu}$. In this gauge the action (1.2)
takes the form
$$S=\int d^2x \sqrt{\bar g}\big[ {1\over 2}\bar g^{\mu\nu}\p_\mu
\varphi\p_\nu\varphi +q\bar g^{\mu\nu}\p_\mu\varphi\p_\nu\gs +
{1\over 2}q\varphi \bar R+V(\varphi)e^{2\gs}\big] \ .
\eqno(2.2) $$
Now the divergences can be computed using the background field method.
It is useful to interpret (2.2) as a particular case of the $D=2$
$\gs $-model
$$S=\int d^2x\sqrt{\bar g}\big[{1\over 2}G_{ij}(X)
\bar g^{\mu\nu}\p_\mu X^i\p_\nu X^j+{1\over 2}\bar R\Psi (X)+T(X)\big]
\ \ , \eqno(2.3)$$
$$X^i=(\varphi,\gs )\ ,\ \ G_{ij}=\left(\matrix
{1&q\cr q&0\cr}\right)\ ,\ \
\Psi =q\varphi\ ,\ \ T= V(\varphi )e^{2\gs}\ .\eqno(2.4)$$
Since the metric $G$ is flat and the dilaton $\Psi $ is linear, the only
non-trivial divergences correspond to a renormalisation of the tachyon.
This is formally true to all orders in the loop expansion. The only
subtle point that may complicate the renormalisation of the model
(1.2) as compared to the $\gs $-model  (2.3),(2.4) is that of
a reparametrization invariance of a cutoff. This issue is irrelevant
in the 1-loop approximation.

The Weyl anomaly coefficient corresponding to the tachyonic
coupling has the following well-known structure [8]:
$$\bar\gb^T=-{1\over 4\pi}G^{ij}D_iD_j T+(G^{ij}\p_i \Psi\p_jT-2T)
\equiv \gb^T+\gD \gb^T \ .\eqno(2.5) $$
Here the first term $\gb^T$ corresponds to the genuine UV divergence
while $\gD\gb^T$ represent the classical Weyl anomaly. Computing
 (2.5) for the particular couplings in (2.4) we find
$$\bar \gb^T=\gb^T= {1\over \pi q^2}e^{2\gs}(V-qV')\ ,
\ \ \gD\gb ^T =0\ .\eqno(2.6)$$
Thus the condition of UV finiteness coincides with the condition
of background Weyl invariance and is satisfied if $V=qV'$, i.e. if
$$V=\mu e^{\varphi/q}\ .\eqno(2.7)$$
It should be stressed that this conclusion holds only if both the
scalar field $and$
the metric are quantised: the scalar models (1.6),(1.8) are not finite
if quantised on a fixed curved background.

Remarkably, the potential (2.7) corresponds to the
``string-inspired" model (1.5) (see eq.(1.8)). It is possible to check
the finiteness of the model (1.5) directly by going into the conformal
gauge in (1.5) and rewriting the resulting action in $\gs $-model form
(1.20) (see (3.18)-(3.21)).
 The UV finiteness condition is satisfied only if a potential
term is the same as in (1.5), i.e. $c \  e^{-2\Phi}\ . $
\footnote{$^5$}{This,
of course, is the expected conclusion since the $\gs$-models
corresponding to (1.5) and (1.8) are related by a field redefinition
which leaves the on-shell finiteness condition  invariant.}
A weaker renormalizability condition is satisfied if $V-qV'$ is proportional
to the potential itself, i.e. in the case of the Liouville potential
$$V=\mu e^{\g\varphi}\ ,\ \ \g={\rm const } \ .\eqno(2.8)$$
 Then the  divergence can be absorbed
into a renormalization of $\mu$ (or a shift of $\varphi $). We conclude
that the Liouville theory is the only 2d scalar field theory
which remains renormalizable after coupling to 2d quantum gravity
(with the linear curvature term included as in (1.2)).

Using the conformal gauge it is easy to compute the
1-loop  effective action
in the theory (1.2). Expanding near the
background $\tilde g_{\mu \nu}=\bar g_{\mu \nu}
\ , \ \ \varphi=\bar \varphi$
and accounting for the mixing between the scalar field and the
conformal factor
 one finds
$$\Gamma_{\rm eff}[\bar g,\bar\vp ]=\half \log\det \gD _{ij}
-\log\det\gD_{\rm gh}\ \  ,\eqno(2.9)$$
where $\gD _{\rm gh}$ is the standard ghost operator corresponding
to the conformal gauge and
$$\gD_{ij}=\gD_{ij}^{(0)}+M_{ij}\ ,\eqno(2.10)$$
$$\gD_{ij}^{(0)}=\left(\matrix{\bar\gD &q\bar\gD \cr
                               q\bar\gD &0\cr}\right)\ ,\ \ \
M_{ij}= \left( \matrix{V''&2V'\cr 2V'&4V\cr}\right)\ ,\ \
\bar\gD=-\bar\nabla ^2\ ,\ \ V'={\p V\over \p \bar\vp } \ . $$
The $V$-independent part of (2.9) is given by
(we fine tune the coefficient of the
 cosmological term $\gl \sqrt{\bar g}$ to zero)
$$\Gamma_{\rm eff}^{(0)}=
{1\over 8}A\int \bar R{\bar \gD}^ {-1}\bar R +{\rm const}\ \  ,
\ \ \ \  A=(-2+26)/12\pi \  \ .\eqno(2.11)$$
The flat
metric scalar Laplacian terms
 $\log\det\gD $ cancel out (the scalar and conformal factor
contributions are compensated by the two ghost contributions)
so that the model has zero dynamical degrees of freedom (diagonalizing
$\gD_{ij}^{(0)}$ one finds that the signs of the eigenmodes are
opposite so that the euclidean path integral should be defined by
 an analytic continuation as in $D=4$ Einstein theory).

In the case of the model (1.8) $V=\mu e^{\vp /q}$ one may compute
(2.9) on the ``black hole" solution (1.15), i.e.
$\bar g_{\mu\nu}=e^{2\bar\gs}\gd _{\mu\nu}\ ,\ \bar\vp=ax^2+b\ $,
$2\bar\gs+\bar\vp/q=k$. Using that $\sqrt{\bar g}V(\bar\vp )
=\mu e^k=\bar\mu={\rm const}$ and that the metric is conformally
flat one finds that the $\bar\mu$-dependent terms in $\det\gD_{ij}$
cancel out so that there is no non-trivial correction to the
anomaly term (2.11) in the effective action (in
 particular, there  are no ``extra" negative modes of $\gD_{ij}$).
If one repeats the calculation by starting with the classical action
in the original parametrization (1.5) one finds again
that the effective action is given simply by (2.11) but
now the background metric $\bar g_{\mu\nu}$ is given by the
classical solution for $g_{\mu\nu}$ (1.16), i.e.
$\bar g_{\mu\nu}={d \ \delta_{\mu\nu}}/(ax^2+b)
$ (we are assuming that starting with (1.5) one defines the
quantum theory in terms of the metric $g_{\mu\nu}$). The resulting
integral in (2.11) is less divergent than in the first case.
If one starts with the action in yet another parametrization
(3.15) and defines the determinants in terms of the metric $\hat g
_{\mu\nu}$ (3.16) (which is flat on the classical solution)
the effective action is given by (2.11) with
$\bar g_{\mu\nu}={\rm const \ } \gd_{\mu\nu}$, i.e. is trivial. Let us
note that starting with (3.15) one can formally argue that the effective
action is trivial to all orders in perturbation theory
(non-perturbatively, however, there is a subtlety related to the fact
that $\psi $ should be non-negative in order to preserve a
correspondence between (1.8) and (3.15)).

Computing (2.9) for the flat metric and constant scalar background
one finds the following expression for the effective potential
$$V_{\rm eff}={1\over 2\pi} \int^{\Lambda^2}_{0}
 d^2 p \ \ln \det (\delta^i_j p^2
+ M^i_j ) \ \ ,$$
$$ M^i_j = G^{ik} M_{kj}\ \ ,\
 \ \ G^{ik}=\left(\matrix{0&-1/q\cr -1/q&-1/q^2 \cr}\right)\ ,$$
\n i.e.
$$V_{\rm eff} = \half m^2_1 (\ln {\Lambda^2\over m^2_1  } + 1 )
+ \half m^2_2 (\ln {\Lambda^2 \over m^2_2} + 1 ) + {\rm const} \ , $$
$$ m^2_{1,2} = 2q^{-2} [ V-qV' \pm ( V^2 -2qVV' + q^2 VV'')^{1/2} ]
\ \ .$$
If $V=\mu e^{\g \varphi} $ then
$ \ \ m^2_1= 4q^{-2} \mu e^{\g \varphi}(1-\g q)\ \ , \ \ m^2_2=0
$
 so that the effective potential vanishes in the case of the
model (1.8) , i.e. when $\g=1/q$.
\bigskip
The usual kinetic term for the conformal factor $\gs $ is missing in
eq.(2.2). Since such term is generated at the 1-loop level
(appearing both from the ghost and scalar determinants)
 one may consider a resummation of the standard
 loop expansion by adding the Weyl anomaly term
$$S_{\rm anom}=
{1\over 8}A\int \tilde R{\tilde \gD}^ {-1}\tilde R
=\half A\int d^2x \sqrt{\bar g}
(\p_\mu\gs\p^\mu \gs + \bar R\gs )
+{1\over 8}A\int \bar R \bar\gD ^{-1}\bar R \ ,\ \eqno(2.12) $$
$$A={1\over 12\pi}(26- D_{\rm eff})\ , $$
 ( $D_{\rm eff}$ is a number of effective scalar degrees of freedom)
to the classical action (2.2) and quantizing the resulting theory.
The corresponding $D=2$ $\gs $-model will be given by eq.(2.3)
with the modified metric and dilaton
$$ G_{ij}=\left(\matrix
{1&q\cr q&A\cr}\right)\ ,\ \  \ \ \Psi =q\varphi +A\gs \ .\eqno(2.13)$$
Computing the ${\bar \gb}^T $-function (2.5) we find
$$\gb ^T=-{1\over 4\pi}{e^{2\gs }\over A-q^2}(AV''-4qV'+4V)\
,\eqno(2.14)$$
$$\gD \gb^T= {e^{2\gs }\over A-q^2}(AqV'-2q^2V-qAV'+2AV)-2Ve^{2\gs}
\equiv 0\ .\eqno(2.15)$$
Eq.(2.14) reproduces (2.6) in the case of $A=0$. As it is clear from
(2.14) the Liouville theory (2.8) is still renormalizable, with the
zero of the $\gb $-function corresponding to the coefficient $\g $ in
(2.8) which satisfies
$$A\g^2-4q\g +4=0 \ \ .\eqno(2.16)$$
We conclude that the model (1.5),(1.8) with the potential (2.7) with
$\g =1/q$ is no longer finite.
The parameter $\mu $ in eq.(1.8),(2.7) (or $c$ in (1.5))
is running with scale unless $A=0$.
 \bigskip
\centerline{\bf 3. Quantum effective action with anomaly}\bigskip
We have seen that the action (1.5) or (1.8) is special being not
renormalised at the quantum level within the naive loop expansion.
It would be interesting to understand its
 modification by finite quantum gravitational corrections and,
in particular, to determine the fate of its classical black hole solution.
As an attempt in this direction we shall
suggest  how one can find a ``quantum" analog of (1.5),(1.8)
 using the DDK - type argument [3].

     In the above discussion in sect. 2  we were ignoring
 the issue of maintaining the general
covariance  of the quantum theory. Instead
of using an invariant regularisation
(which is  complicated in the conformal gauge) one may adopt
 a non-invariant cutoff adding at the same time some counterterms
which are necessary in order to satisfy the reparametrisation invariance
Ward identities. The resulting ``effective action" should generate
a theory which is invariant under the background
Weyl symmetry [3],
$\bar g\to e^{2\tau(x)}\bar g , \gs (x)\to\gs (x)-\tau (x)$.
Since the metric $\tilde g= e^{2\gs} \bar g$ is left unchanged
 this transformation
should be an exact symmetry of the theory, i.e. the $\bar \gb $-functions
of the couplings in the ``effective action" should vanish.
  The basic assumption (which can be justified
to a certain extent in the absence of a scalar potential)
 is that the conformal factor
dependence of the covariant quantum measure and regularisation
can be represented by a {\it local}  ``effective action"
containing only simplest lowest derivative terms [3].

Since in the conformal gauge the actions
(1.5),(1.8)  describe the systems of $two$
interacting scalar fields a priori one might expect that the corresponding
``effective action" is a general $D=2$ $\gs$-model  with
the couplings being
 functions of the two arguments which solve the Weyl
invariance conditions. It is natural to assume, however,
 that for a particular choice of the  quantum variables
 only some particular structures are actually appearing
In what follows we shall discuss two (inequivalent)
 suggestions for the  effective action which may correspond
to (1.5),(1.8).  They will be described by two particular  $D=2$
 $\gs$-models  which generalize the actions  (1.5) and (1.8)
to the case when  the anomaly terms  are included.

While equivalent at the classical level (being
related by the field redefinition (1.7))  the models (1.5) and (1.8)
 are not necessarily equivalent
at the quantum level since the field
 redefinition involves a Weyl rescaling of the metric
which may introduce an additional anomaly.
 Given that  the scalar field and the conformal factor
are mixed in the classical action it is not
obvious which particular metric should be used in
 the
definition of the quantum theory (i.e. in  a measure and in a cutoff).
\footnote{$^6$}{A ``preferred" metric may be selected by coupling a
gravitational system to extra matter fields.}
\bigskip
Let us first discuss the ansatz for the effective action which
is  most natural if we use  (1.8) as a starting point.
We shall assume that the clasical action (2.2) should be replaced at the
quantum level by (cf.(2.2),(2.12))
$$S_{\rm  eff}
 =\int d^2x\sqrt{\bar g}\big[{1\over 2}\bar g^{\mu\nu}\p_\mu
\varphi\p_\nu\varphi +{1\over 2}A\bar g^{\mu\nu}\p_\mu\gs\p_\nu\gs
+q\bar g^{\mu\nu}\p_\mu\varphi\p_\nu\gs $$
$$+{1\over 2}(q\varphi +B\gs )\bar R+ T(\varphi,\gs ) \big]\ .
\eqno(3.1)$$
Since the two scalar fields can be linearly combined
and rescaled  in order to eliminate the mixing term the only
a priori free parameters  (in addition to $T$) are the two
coefficients of the curvature couplings
(the coefficient of the mixing term can be taken  to be
equal to  $q$ without loss of generality).
We shall fix $A$ at  its natural value
 $A={25-D_{\rm eff}\over 12\pi }$,  where
$D_{\rm eff} =1+N$ and $N$ is a number of extra matter
scalar fields which may contribute to the conformal anomaly.
The condition of the vanishing of the total central
charge gives (see (2.3),(2.13))
$$\bar \gb^{\Psi}={1\over 12\pi }(D_{\rm eff} +1-26)+
G^{ij} \p_i \Psi \p_j \Psi$$ $$={1\over 12\pi} (D_{\rm eff}-25)+
{1\over A-q^2}[Aq^2-2q^2B+B^2]=0       \ \ . \
\eqno(3.2)$$
 Eq.(3.2) has  two solutions:
$B=2q^2-A$ and $B=A$. The correspondence with the classical
limit $A=0$ selects  the solution $B=A$.
A solution of the condition of the
vanishing of the ``tachyon" Weyl anomaly  coefficient (2.5) is given by
 $T=\mu e^{a\gs+\g \varphi}$, where  the constants
$a$ and $\g $  satisfy the relation
$$-{1\over 2}A \g ^2-{1\over 2}a^2+ q\g a+
2\pi (a-2)(A  - q^2 ) =0    \ .
\eqno(3.3)$$
Eq.(3.3) is not sufficient in order
 to determine both $ a$ and $\g$. We
shall make an additional assumption
of correspondence between the ``effective action" and (1.8),(1.5)
in the classical limit. In particular,
the effective equations should have asymptotically flat solutions.
Let us solve the vacuum
equations for arbitrary $a$ and $\g $.
Setting $\bar g_{\mu\nu}=\gd_{\mu\nu}$ and introducing the complex
coordinates $z, \bar z$ we find that
the  equations of motion which follow from
 (3.1) become
$$\p\bar\p  \varphi+q\p\bar\p \gs={\mu\over 4} \g e^{a\gs +
\g \varphi}\ , \eqno(3.4)$$
$$q\p\bar\p\varphi +A\p\bar\p\gs ={\mu\over 4} a e^{a\gs +\g\varphi}
 \ . \eqno(3.5)$$
Combining eqs.(3.4),(3.5) we  get
$$\gs =- \ga \varphi \ ,\ \ \
\ga\equiv {a-\g q\over aq-A\g } \ \eqno(3.6)$$
(we have dropped the sum of
 arbitrary analytic and anti-analytic functions which may appear in
 $\gs$). Inserting  $\gs$ (3.6) into eq.(3.4) we obtain
$$\p\bar\p \varphi =\bar\mu e^{(\g -a\ga )\varphi}\ ,
\ \ \ \bar\mu \equiv{\mu\g \over 4(1-q\ga )}\ . \  \eqno(3.7)$$
 Eq.(3.7) is the  Liouville equation
with the general solution
$$\varphi ={1\over a\ga-\g }\log\bigg[ {\bar\mu (a\ga -\g )
(1+f(z)\bar f(\bar z))
^2\over 2 f'(z)\bar f'(\bar z)}\bigg]\ \ .  \eqno(3.8)$$
One can check that this solution  satisfies also the constraints
implied by general covariance.
This solution does not approach the classical solution (1.15),(1.16)
 unless
$a\ga =\g $. In the latter  case an important simplification
occurs and the solution of (3.7) takes the form
$$\varphi ={\mu\g a\over 4(a-q\g)} z\bar z +{\rm const} \ .
\eqno(3.9)$$
Using the definition of $\ga $ in (3.6) we conclude that
 $a\ga =\g $ implies
the following relation between $\g $ and $a $:
$$1-2q{\g \over a}+A{\g ^2\over a^2}=0 \ . \eqno(3.10)$$
Combining eq.(3.10) with eq.(3.3) we obtain
$$a=2\ , \ \ \
\g ={2q\over A}(1-C)\ ,\ \ \
\ C\equiv (1-{A\over q^2})^{1/2} \ . \eqno(3.11)$$
In the limit $A\to 0$ the value $\g =1/q$ is reproduced.

The transformation which connects $(\varphi ,\tilde g)$ and $(\Phi ,g)$
should be such that
the parameter $q$ disappears in the effective action for $\Phi $ and
$g$ since it was absent in the classical action (1.5) ($q$ can be
eliminated by shifting $\Phi$ ).
 The required
transformation for $A\neq 0 $ is (cf.(1.7))
$$a\gr =\g \varphi - \log({Cq\varphi / 2})\ ,\ \ \
 qC\varphi ={1\over 4}e^{-2\Phi}\ \ ,\ \ \gl= \gr + \gs \
  \ , \ \eqno(3.12)$$
where $\gl $ is the conformal factor of metric $g_{\mu\nu}\ , $ i.e.
$$g_{\mu\nu}=e^{2\gr}\tilde g_{\mu \nu}=e^{2\gl}\bar g_{\mu\nu}\ .$$
Inserting (3.12) into eq.(3.1)  and assuming that $a=2$ we find that the
resulting effective action in terms of the original
variables  $\ (g,\ \Phi)$  of (1.5) is
$$\SE=\int d^2x \sqrt{\bar g}\big[\ {1\over 8} e^{-2\Phi}\big(\bar R +
4\pn\Phi\pmu\Phi-4\pn\gl\pmu\Phi+c\ e^{2\gl}\big)  $$
$$+ A (\  {1\over 2}\pn\gl\pmu\gl+{1\over 2}\pn\Phi\pmu\Phi-\pn\Phi
\pmu\gl+{1\over 2}(\gl-\Phi)\bar R \ ) \big]\ . \eqno(3.13) $$
By construction this action should give the
 vanishing Weyl anomaly coefficients.
This can be checked explicitly and made more transparent
 by rewriting (3.13) in terms of
the new variables $$\psi={1\over 4}e^{-2\Phi}\ \ \ , \ \ \ \k=\gl-\Phi \ \ , $$
$$ \SE= \int d^2x \sqrt{\bar g}\big[\ {1\over 2} (
\bar R \psi + 2 \pn\psi\pmu\k +{1\over 4} c \ e^{2\k} \ )
+\half A (\pn\k\pmu\k + \bar R \k) \big] \ \ , \eqno(3.14)$$
i.e.
$$ \SE= S + S_{\rm anom}\ , \ \ \ S=
{1\over 2} \int d^2x\sqrt{\hat g}\big( \hat R \psi
+{1\over 4} c\ \big) \ \ , \eqno(3.15)$$
$$S_{\rm anom}= \ {1\over 8}A\big[\int \hat R {\hat \Delta}^{-1}
\hat R  \ - \ \int \bar R {\bar \Delta}^{-1} \bar R \ \big]
\ , $$
where we have introduced the metric
$$ \hat g_{\mu \nu} = e^{2\k}\bar g_{\mu \nu} =
 e^{-2\Phi} g_{\mu \nu} = 4\psi g_{\mu\nu} \ \ . \eqno(3.16)$$

The classical solution of (1.5), i.e. of (3.15) with $A=0$
corresponds to the flat $\hat g_{\mu \nu}\ $, i.e. $\hat R =0$.
The structure of (3.15) (namely, the absence of $\psi $ in
its quantum part $S_{\rm anom}$)
 implies that the same is true also
for a non-vanishing $A$, i.e. the extremum of (3.15) is given by
\footnote{$^7$}{ This solution can be  obtained  also
from (3.6), (3.9) by using the transformation
(3.12).}
$$\k=\gl-\Phi=0 \ ,\ \ \ \ \ e^{-2\Phi}= {c\over 4}z\bar z
+{M\over \sqrt c}\ \ .\eqno(3.17)$$
This is  still the classical black hole solution, i.e.
the incorporation of the
quantum Weyl anomaly according to (3.13),(3.15)
has not altered the classical result.

 As it is clear from (3.15) the classical action (1.5)
 has a  very simple representation  in terms of
$\hat g $ and $\psi$.
\footnote{$^8$}
{This representation makes perturbative finiteness of the model
 (1.5) manifest and also simplifies the analysis of the classical solutions
in the case of a non-trivial scalar potential (1.18), i.e. when
$$S= \half \int d^2 x \big[ \hat R \psi + {1\over 4} U(\psi) \big]\ \ ,
\ \ U=c + c_1 \psi^{-1} + c_2 \psi^{-2} + ... \ \ . $$ }
 It appears as if $\hat g $ is a
 ``preferred" metric in terms of which the anomaly
contribution is to be constructed.
 One could have found (3.15)
by directly supplementing the ``naive" conformal anomaly term
$\int \pn \gl\pmu\gl $ by extra terms needed to satisfy the
condition of the background Weyl invariance.
\bigskip
The action (3.15) is not, however, a unique quantum extension of (1.5)
which includes the anomaly term and satisfies the condition of
background Weyl invariance. To find another one  let us start
directly with the action (1.5)
$$ S=\half \int d^2x\sqrt{g}
[\ \ps \pn\psi\pmu\psi + \psi R + c\ \psi \  ]
\ \ , \ \ \ \psi= {1\over 4 }e^{-2\Phi}\ , \eqno(3.18)$$
written in the conformal gauge $g_{\mu\nu}=e^{2\gl}
{\bar g}_{\mu\nu}$ (cf. (3.13))
 $$ S = \half \int d^2x \sqrt{\bar g}
 [  \ps \pn\psi\pmu\psi + 2\pn\psi\pmu\gl
+  \psi\bar R + c\ \psi e^{2\gl}\ ] \ \ . \eqno(3.19)$$
Let us assume that the anomaly term to be added to (3.19)
has its standard $\gl$-dependent form.
The suggested ansatz for the quantum
analog of (3.19) is thus the following
$$ \SE'=\half \int d^2x \sqrt{\bar g}
[\ \ps\pn\psi\pmu\psi + 2\pn\psi\pmu\gl+  A
\pn\gl\pmu\gl +  \bar R (\psi + B\gl )+ T(\psi,\gl)\ ] \ \ ,
\ \eqno(3.20)$$
where $B$ and  $T$ are to be determined from the
background Weyl invariance condition.
It is straightforward to check that the $\gs$-model which
corresponds to (3.20), i.e. (2.3) with
$$ X^i=(\psi,\gl)\ ,\ \ \ \  G_{ij}=\left(\matrix{\ps &1\cr 1&A\cr}
\right) \ ,\ \ \ \Psi= \psi + B\gl \ , \eqno(3.21)$$
has the vanishing metric Weyl anomaly coefficient ${\bar \gb}^G$
provided $B=A$. In fact,  the metric in (3.21)
is actually flat (for any $A$) and the second covariant
derivative of the dilaton $\Psi$
vanishes if $B=A$.
The condition of the vanishing of the total central charge
(3.2) is  satisfied if $A= {1\over 12\pi} ( 25-D_{\rm eff})$.
The vanishing of the tachyon Weyl anomaly coefficient
 ${\bar\gb}^T$  (2.5) gives
$$A\p_\psi^2T-2\p_\psi\p_\gl T+\ps\p_\gl^2 T+{A\ps\over 2(A-\psi)}
(A\p_\psi -\p_\gl) T$$
$$- 4\pi \ps (A-\psi)(\p_\gl -2)T=0  \ . \eqno(3.22)$$
Remarkably, there is no solution to
eq.(3.22) of the ``classical" type $\psi f(\gl )$ and hence the
scalar field dependence of the potential should change.
 There are solutions of the form
$T=e^{a\gl}W(\psi)$. However, if we want $W$ to approach
 $c\psi$ in the classical
region $\psi\to \infty$ we may set $a=2$
(this is similar to the $a=2$ condition (3.11) we have discussed
 above in the derivation of (3.13)). We are thus
assuming that only the $\psi$ dependence of the potential gets modified.
 Then (3.22) gives the following differential equation
for the potential $W$
$$A\psi(A-\psi )W''+[\half A^2-4\psi(A-\psi)]W'+(3A-4\psi)W=0\ \ ,
\eqno(3.23)$$
$$ T=e^{2\gl} W(\psi) \  .$$
In the classical case $A=0 \ ,  \ $  $W=c\ \psi$.
For a general $A$ one can find a solution of (3.23)  by expanding
$$W=c\psi+c_1+c_2\ps+O(\psi ^{-2})\ .\eqno(3.24)$$
Eq.(3.23) gives
$$c_1=-{A\over 4}c\ \ ,\ \ \ c_2=-{A^2\over 32}c\ \  ,\ \ \ {\rm etc.}
\eqno(3.25)
$$
Eq.(3.24) is analogous to the expansion of a string loop corrected
dilaton potential
in terms of a string coupling (cf. eq.(1.18)). If eq.(3.23) were true in
string theory it would give a nonperturbative expression for
the dilaton potential. In sect. 2 we have seen that the presence
of the ``string loop corrections" in the potential (1.18)
 changes the classical black hole geometry
(see eq.(1.26)).
 It is interesting to note that in the case of
the effective  action (3.20)
the  combined effect of the anomaly
term $A\p\gl\p\gl$ and the first correction
$c_1$  to the potential (3.24) is such that
there is no substantial modification
of the black hole solution in the asymptotic weak coupling
 region. A solution of the  equations corresponding to (3.20)
with a general $T$ satisfying (3.23) is likely to be very different
from the classical ($A=0$) solution.

\bigskip
Comparing the two  suggestions for the effective action (3.14) and (3.20)
it is clear that they correspond to the two possible choices of a
metric in terms of which the anomaly contribution is constructed.
In (3.14) this is $\hat g$ (3.16) which is natural from the point of
view of the ``canonical" form of the
 action (1.8). The use of $\hat g$ makes it possible
to keep the classical form of the potential term and of the
vacuum solution. In (3.20) one employs the original
metric $g$ of (1.5) but then the potential term and  the
vacuum solution become complicated.

In ref.[2]  the effect
of  quantum corrections on the  black hole
solution in the  model (1.5) was analysed in the $1/N$
approximation (i.e. the metric and
 $\phi $ were considered as classical). The
inclusion of the conformal anomaly term
due to scalar matter fields has led to a  drastic
modification of the classical   solution.
Our discussion suggests that once the quantum gravitational
corrections are accounted for in a way consistent
with  the general covariance (so that the ``effective action"
in the conformal gauge satisfies the
condition of the background Weyl invariance) the structure of
the anomaly terms  and hence of the corresponding vacuum solutions
may be different.
To justify further the ansatze (3.15) or (3.20) it is important to
understand  a relation
between an approximation in which they can be considered as
candidates for a quantum effective action
and the standard  loop and $1/N$ expansions.
\bigskip
\bigskip
A.T. would like to acknowledge a financial
support of Trinity College, Cambridge. The research of J.R.
is supported by INFN.
\vfill\eject
\centerline {\bf References}\bigskip

\item{[1]}E. Witten, Phys.Rev. D44 (1991) 314;

  S. Elitzur, A. Forge and E. Rabinovici, Nucl.Phys. B359 (1991) 581;

G. Mandal, A. Sengupta and S. Wadia, Mod.Phys.Lett. A6 (1991) 1685.

\item{[2]}C.G. Callan, S.B. Giddings, J.A. Harvey and
A. Strominger,

Santa Barbara preprint UCSB-TH-91-54.

\item{[3]}F. David, Mod.Phys.Lett. A3 (1988) 1651;
Nucl.Phys. B293 (1988) 332;

     J. Distler and H. Kawai, Nucl.Phys. B321 (1989) 509;

     S.R. Das, S. Nair and S.R. Wadia, Mod.Phys.Lett. A4 (1989) 1033;

     J. Polchinski, Nucl.Phys. B324 (1989)123;

     T. Banks and J. Lykken, Nucl.Phys. B331 (1990) 173;

     A.A. Tseytlin, Int.J.Mod.Phys. A5 (1990) 1833;

     E. D'Hoker, Mod.Phys.Lett. A6 (1991) 745;

A. Cooper, L. Susskind and L. Thorlacius, Nucl.Phys. B363 (1991) 132.

\item{[4]}R. Marnelius, Nucl.Phys. B211 (1983) 14;

  C. Teitelboim, Phys.Lett.Phys.Lett. B126 (1983) 41;

  T. Yoneya, Phys.Lett. B148 (1984) 111;

 R. Jackiw, in: Quantum Theory of Gravity, ed. S.Christensen (Adam

Hilger, Bristol 1984);

 A. Chamseddine, Phys.Lett. B256 (1991) 2930.

\item{[5]}T. Banks and M. O'Loughlin, Nucl.Phys. B362 (1991) 649.

\item{[6]}M. McGuigan, C. Nappi and S. Yost, IAS preprint IASSNS-HEP-91-57.

\item{[7]}S.D. Odintsov and I.L. Shapiro, Phys.Lett. B263 (1991) 183;

Madrid preprint FTUAM-33 (1991);

   D. Mazzitelli and N. Mohammedi, Trieste preprint IC/91/238 (1991).

\item{[8]} C.G. Callan, D. Friedan, E. Martinec and M.J. Perry,

     Nucl.Phys. B262 (1985) 593.

\vfill\eject
\end